\documentclass[aps,pra,twocolumn,floatfix,showpacs,reprint,superscriptaddress]{revtex4-1}

\usepackage{graphicx}
\usepackage{amsmath, amsfonts, amssymb, bm}

\usepackage{amstext}
\usepackage{mathrsfs}
\usepackage{color}
\newcommand{\corr}[1]{{\color{black}#1}}
\newcommand{\cor}[1]{{\color{black}#1}}

\begin{document}
\title{\corr{Implementing nonlinear Compton scattering beyond}\\
\corr{the local constant field approximation}}
\author{A.~Di Piazza}
\email{dipiazza@mpi-hd.mpg.de}
\affiliation{Max-Planck-Institut f\"ur Kernphysik, Saupfercheckweg 1, D-69117 Heidelberg, Germany}
\author{M.~Tamburini}
\email{mtambu@mpi-hd.mpg.de}
\affiliation{Max-Planck-Institut f\"ur Kernphysik, Saupfercheckweg 1, D-69117 Heidelberg, Germany}
\author{S.~Meuren}
\email{smeuren@princeton.edu}
\affiliation{Department of Astrophysical Sciences, Princeton University, Princeton, New Jersey 08544, USA}
\affiliation{Max-Planck-Institut f\"ur Kernphysik, Saupfercheckweg 1, D-69117 Heidelberg, Germany}
\author{C.~H.~Keitel}
\email{keitel@mpi-hd.mpg.de}
\affiliation{Max-Planck-Institut f\"ur Kernphysik, Saupfercheckweg 1, D-69117 Heidelberg, Germany}

\begin{abstract}
In the calculation of probabilities of physical processes occurring in a 
background classical field, the local constant field approximation (LCFA) 
relies on the possibility of neglecting the space-time variation of the 
external field within the region of formation of the process. This approximation 
is widely employed in strong-field QED as it allows to evaluate probabilities 
of processes occurring in arbitrary electromagnetic fields starting from the 
corresponding quantities computed in a constant electromagnetic field. Here, 
we \cor{scrutinize the validity of the LCFA in the case of nonlinear
Compton scattering focusing on the role played by the energy of the emitted
photon on the formation length of this process. In particular, we derive 
analytically the asymptotic behavior of the emission probability per unit of 
photon light-cone energy $k_-$ and show that it tends to a constant for $k_-\to 0$. 
With numerical codes being an essential tool for the interpretation
of present and upcoming experiments in strong-field QED, we obtained an 
improved approximation for the photon emission probability, implemented it 
numerically, and showed that it amends the inaccurate behavior of the LCFA 
in the infrared region, where it remains in qualitative and good quantitative 
agreement with the full strong-field QED probability.}
\end{abstract}

\maketitle

\section{Introduction}
\cor{QED is probably the best tested and most successful among the physical theories. There are, however, sectors of QED which still need to be thoroughly scrutinized both theoretically and experimentally. In particular, the so-called strong-field sector of QED has recently attracted considerable attention. The strong-field sector of QED includes electrodynamical processes occurring in the presence of background electromagnetic fields of the order of the ``critical'' field scale: $F_{cr}=m^2/|e|=1.3\times 10^{16}\;\text{V/cm}$ \cite{Landau_b_4_1982,Fradkin_b_1991,Dittrich_b_1985}. Here, $m$ and $e<0$ are the electron mass and charge, respectively, and units with $\hbar=c=4\pi\epsilon_0=1$ and $\alpha=e^2\approx 1/137$ are employed throughout the paper.} High-power lasers are a unique tool to test QED in the strong-field sector, whose field scale corresponds to an intensity $I_{cr}=F_{cr}^2/4\pi=4.6\times 10^{29}\;\text{W/cm$^2$}$. The field strength $F$ of present and soon available lasers is still much smaller than $F_{cr}$ \cite{Yanovsky_2008,APOLLON_10P,ELI,XCELS}. \cor{However, probing QED at the critical field scale is facilitated by the Lorentz invariance of the theory. The latter implies that the parameters controlling QED processes must be Lorentz-invariant quantities. Thus, the effective field scale at which a process occurs is not set by the field amplitude $F$ in the laboratory frame but rather} by the amplitude $F^*$ that the participating charged particles experience in their rest frame \cite{Mitter_1975,Ritus_1985,Ehlotzky_2009,Reiss_2009,Di_Piazza_2012,Dunne_2014}. Indeed, the parameter which identifies the strong-field QED regime is given by $\chi_0=F^*/F_{cr}$ \cite{Mitter_1975,Ritus_1985,Ehlotzky_2009,Reiss_2009,Di_Piazza_2012,Dunne_2014}. Available technology already allows for entering the strong-field QED regime ($\chi_0\gtrsim 1$) by combining either conventional \cite{Bula_1996,Burke_1997} or laser-based \cite{Leemans_2014} multi-GeV electron accelerators with high-power optical lasers \cite{Yanovsky_2008,APOLLON_10P,ELI}. Another promising setup is based on the interaction of an ultra-intense laser beam with a solid target \cite{Ridgers_2012,Bulanov_2013,Nerush_2014,Brady_2014,Chang_2015,Gong_2017}.

Strong-field QED processes in intense laser beams are conveniently studied theoretically within the plane-wave approximation, as the resulting Dirac equation can be solved analytically \cite{Landau_b_4_1982,Fradkin_b_1991,Dittrich_b_1985}. Now, ultrarelativistic charges are typically involved in considered strong-field QED processes, such that an arbitrary external electromagnetic field looks as a plane wave in the instantaneous rest frame of the charge \cite{Jackson_b_1975}. In addition, since the background laser fields considered in applications are typically very intense, the basic strong-field QED processes (nonlinear Compton scattering and nonlinear Breit-Wheeler pair production) are formed on a length scale much smaller than the laser wavelength \cite{Ritus_1985,Di_Piazza_2012}, and the corresponding available probabilities in the so-called local constant field approximation (LCFA) can be employed. 

Generally speaking, the LCFA is said to be applicable when the background laser field is so strong that $\xi_0=|e|E_0/m\omega_0\gg 1$ \cite{Ritus_1985,Di_Piazza_2012}, where $E_0$ is the laser field amplitude and $\omega_0$ its central angular frequency. Since at optical laser photon energies ($\omega_0\sim 1\;\text{eV}$), the condition $\xi_0\sim 1$ is already satisfied at intensities of the order of $10^{18}\;\text{W/cm$^2$}$, the probabilities of the basic QED processes within the LCFA are widely employed to interpret present experiments on strong-field QED at intensities above $10^{20}\;\text{W/cm$^2$}$, to predict the results of upcoming ones, and to investigate theoretically QED effects in laser-plasma interaction. More precisely, the condition of validity of the LCFA in the quantum regime $\chi_0\gtrsim 1$ was found to be  $\xi_0^3/\chi_0\gg 1$ \cite{Baier_1989,Dinu_2016}. However, at, e.g., $\xi_0\gtrsim 10$, the violation of the additional condition $\xi_0^3/\chi_0\gg 1$ implies that $\alpha\chi_0^{2/3}\gtrsim 1$ and then that the perturbative approach to strong-field QED in a plane wave breaks down \cite{Ritus_1985}.

Here, we first challenge the validity of the LCFA at $\xi_0\gg 1$ by investigating analytically the photon emission probability in nonlinear single Compton scattering. We find that even for a plane wave with $\xi_0\gg 1$, the LCFA predicts a photon yield which differs quantitatively and qualitatively from the exact one for $k_-\lesssim (\chi_0/\xi_0^3)p_-$, with $k_-$ ($p_-$) being the photon (electron) light-cone energy \footnote{The light-cone energy of a particle is the difference between  the particle's energy and its component of the momentum along the plane-wave propagation direction.}. The origin of the failure of the LCFA here is the dependence of the formation length of the process on the photon light-cone energy (see, e.g., \cite{Baier_1989,Blankenbecler_1996}) rather than the interference effects investigated in \cite{Harvey_2015,Meuren_2016}. Then, we provide an improved approximation of the differential emission probability, that can be straightforwardly implemented in numerical codes, which are routinely employed to analyze and interpret experiments in strong-field QED. Unlike the differential probability within the LCFA, the improved expression is finite at low photon light-cone energies and this difference is shown to be significant and in principle already observable experimentally. 

Note that the leading-order correction to the LCFA, which depends on the derivatives of the background field, has been derived analytically in \cite{Baier_1989}, which, however, introduces a fictitious non-integrable divergence in the infrared limit of the emission probability (the divergence in the intensity of radiation is integrable, though). Furthermore, strong deviations from the LCFA have been predicted in the low-energy part of the power spectrum emitted by multi-GeV electrons in oriented crystals \cite{Khokonov_2002}. Analogous deviations as in \cite{Khokonov_2002} have been predicted in \cite{Wistisen_2015} in the case of a uniform magnetic field of finite extension. Here we investigate the case of nonlinear Compton scattering and we provide the first numerical implementation of this process beyond the LCFA, that employs only the local value of the background field at the interaction point. This is an important advantage, as it significantly reduces the computational overhead required to improve the LCFA.

\section{Theoretical Model}
Below we consider an electron with incoming four-momentum $p^{\mu}=(\varepsilon,\bm{p})$, which collides with a plane wave propagating along the $\bm{n}$ direction ($\bm{n}^2=1$). The plane wave is characterized by the four-vector potential $A^{\mu}(\phi)=(0,\bm{A}_{\perp}(\phi))$, where $\phi=(nx)=t-\bm{n}\cdot\bm{x}$ [i.e., $n^{\mu}=(1,\bm{n})$], and where $\bm{n}\cdot\bm{A}_{\perp}(\phi)=0$ and $\lim_{\phi\to\pm\infty}\bm{A}_{\perp}(\phi)=\bm{0}$  [i.e., the four-potential is chosen in the Lorentz gauge $\partial_{\mu}A^{\mu}(\phi)=0$]. Several aspects of nonlinear single Compton scattering have been already studied recently \cite{Ivanov_2004,Boca_2009,Harvey_2009,Mackenroth_2010,Boca_2011,Mackenroth_2011,Seipt_2011,Seipt_2011b,Krajewska_2012,Seipt_2013,Krajewska_2014,Wistisen_2014,Harvey_2015,Seipt_2016,Seipt_2016b,Angioi_2016,Harvey_2016b} (see also the reviews \cite{Mitter_1975,Ritus_1985,Ehlotzky_2009,Reiss_2009,Di_Piazza_2012}). In particular, since the plane wave depends only on the variable $\phi$, it is clearly convenient to introduce the light-cone coordinates $T=(t+\bm{n}\cdot\bm{x})/2$, $\bm{x}_{\perp}=\bm{x}-(\bm{n}\cdot\bm{x})\bm{n}$, and, indeed, $\phi=t-\bm{n}\cdot\bm{x}$, as well as the light-cone components $v_+=(v^0+\bm{n}\cdot\bm{v})/2$, $\bm{v}_{\perp}=\bm{v}-(\bm{n}\cdot\bm{v})\bm{n}$, and $v_-=v^0-\bm{n}\cdot\bm{v}$ of an arbitrary four-vector $v^{\mu}=(v^0,\bm{v})$ (note that $T=x_+$ and $\phi=x_-$). Assuming that the emitted photon (outgoing electron) is characterized by a four-momentum $k^{\mu}=(\omega,\bm{k})$ [$p^{\prime\mu}=(\varepsilon',\bm{p}')$], the leading-order emission probability $dP/dk_-$ averaged (summed) over all initial (final) discrete quantum numbers has been derived in detail, e.g., in \cite{Di_Piazza_2016,Di_Piazza_2017b} and, for the sake of completeness, some technical details are summarized below. The leading-order matrix element of the process within the Furry picture \cite{Landau_b_4_1982} is
\begin{equation}
\label{S_C}
S_{fi}=-ie\sqrt{4\pi}\int d^4x\,\bar{\psi}_{p',s'}(x)\frac{\hat{e}_{k,l}}{\sqrt{2\omega V}}e^{i(kx)}\psi_{p,s}(x),
\end{equation}
where $\psi_{p,s}(x)$ ($\psi_{p',s'}(x)$) is the positive-energy Volkov state with asymptotic four-momentum $p^{\mu}$ ($p^{\prime\,\mu}$) and asymptotic spin quantum number $s$ ($s'$), i.e., the positive-energy solution of the Dirac equation in the plane wave $A^{\mu}(\phi)$ \cite{Landau_b_4_1982}, where the hat on a four-vector indicates the contraction of the latter with the Dirac gamma matrices $\gamma^{\mu}$, where in general $\bar{\psi}=\psi^{\dag}\gamma^0$ for an arbitrary bispinor $\psi$, where $e^{\mu}_{k,l}$ is the (linear) polarization four-vector of the emitted photon, and where $V$ is the quantization volume. For the sake of completeness, we report here the expression of the positive-energy Volkov states \cite{Landau_b_4_1982}:
\begin{align}
\label{Volkov}
\psi_{p,s}(x)=&e^{iS_p(x)}\left[1+\frac{e}{2 p_-}\hat{n}\hat{A}(\phi)\right] \frac{u_{p,s}}{\sqrt{2V\varepsilon}},\\
S_p(x)=&-(px)-\int_0^{\phi}d\phi'\left[\frac{e(pA(\phi'))}{p_-}-\frac{e^2A^2(\phi')}{2p_-}\right],
\end{align}
where $u_{p,s}$ is the constant bispinor solution of the equation $(\hat{p}-m)u_{p,s}=0$. Since the external field depends only on the variable $\phi$, one is able to carry out the three integrations over $T$ and $\bm{x}_{\perp}$ analytically, and one obtains
\begin{widetext}
\begin{equation}
\label{S_C_2}
\begin{split}
S_{fi}=&-ie\sqrt{\frac{\pi}{2V^3\varepsilon\varepsilon'\omega}}(2\pi)^3\delta^{(2)}(\bm{p}'_{\perp}+\bm{k}_{\perp}-\bm{p}_{\perp})\delta(p'_-+k_--p_-)\int d\phi\,\bar{u}_{p',s'}\left[1-e\frac{\hat{n}\hat{A}(\phi)}{2p'_-}\right]\hat{e}_{k,l}\left[1+e\frac{\hat{n}\hat{A}(\phi)}{2p_-}\right]u_{p,s}\\
&\times\exp\left\langle i\left\{(p'_++k_+-p_+)\phi+\int_0^{\phi}d\phi'\,\left[e\frac{(p'A(\phi'))}{p'_-}-e\frac{(pA(\phi'))}{p_-}-e^2\frac{A^2(\phi')}{2}\left(\frac{1}{p'_-}-\frac{1}{p_-}\right)\right]\right\}\right\rangle.
\end{split}
\end{equation}
\end{widetext}
The average probability $dP$ that a photon is emitted with momentum between $\bm{k}$ and $\bm{k}+d\bm{k}$ is given by
\begin{equation}
dP=V\frac{d^3\bm{k}}{(2\pi)^3} V\int\frac{d^3\bm{p}'}{(2\pi)^3} \frac{1}{2}\sum_{l,s,s'}|S_{fi}|^2.
\end{equation}
When squaring modulus of the matrix element, one only has to take care of the fact that the square of the $\delta$-function $\delta(p'_-+k_--p_-)$ is performed after transforming it as $\delta(p'_-+k_--p_-)=(\varepsilon/p_-)\delta(p_n-p_n^*)$, where $p_n=\bm{p}\cdot\bm{n}$ and $p_n^*=(m^2+\bm{p}_{\perp}^2-P_-^2)/2P_-$, with $P_-=p'_-+k_-$, in such a way that the usual procedure exploiting the periodic boundary conditions along the $\bm{n}$ direction can be exploited. Moreover, the sum over the spin variables ($\sum_s u_{p,s}\bar{u}_{p,s}=\hat{p}+m$ and $\sum_{s'} u_{p',s'}\bar{u}_{p',s'}=\hat{p}'+m$) and over the photon polarization ($\sum_le_{k,l}^{\mu}e_{k,l}^{\nu}\to-g^{\mu\nu}$, where the arrow recalls that the substitution is allowed by gauge invariance, see \cite{Landau_b_4_1982}) leads to appearance of the trace:
\begin{widetext}
\begin{equation}
\mathcal{T}=-\frac{1}{4}\text{Tr}\left\{(\hat{p}'+m)\bigg[1-\frac{e}{2 p'_-}\hat{n}\hat{A}(\phi)\bigg]
\gamma^{\mu}\bigg[1+\frac{e}{2 p_-}\hat{n}\hat{A}(\phi)\bigg](\hat{p}+m)\bigg[1-\frac{e}{2 p_-}\hat{n}\hat{A}(\phi')\bigg]\gamma_{\mu}\bigg[1+\frac{e}{2 p'_-}\hat{n}\hat{A}(\phi')\bigg]\right\}.
\end{equation}
\end{widetext}
The evaluation of $\mathcal{T}$ can be carried out with the standard technique as explained, e.g., in \cite{Landau_b_4_1982} and the result is
\begin{widetext}
\begin{equation}
\begin{split}
\mathcal{T}=&m^2\left(\frac{p'_-}{p_-}+\frac{p_-}{p'_-}-4\right)+\frac{p'_-}{p_-}\bm{p}^2_{\perp}-2\bm{p}_{\perp}\cdot\bm{p}'_{\perp}+\frac{p_-}{p'_-}\bm{p}^{\prime\, 2}_{\perp}+ek_-\left(\frac{\bm{p}_{\perp}}{p_-}-\frac{\bm{p}'_{\perp}}{p'_-}\right)\cdot[\bm{A}_{\perp}(\phi)+\bm{A}_{\perp}(\phi')]\\
&-e^2\left[\bm{A}^2_{\perp}(\phi)+\bm{A}^2_{\perp}(\phi')-\left(\frac{p'_-}{p_-}+\frac{p_-}{p'_-}\right)\bm{A}_{\perp}(\phi)\cdot\bm{A}_{\perp}(\phi')\right],
\end{split}
\end{equation}
\end{widetext}
where the conservation laws imply that $\bm{p}'_{\perp}=\bm{p}_{\perp}-\bm{k}_{\perp}$ and that $p'_-=p_--k_-$. By using the above expression of $\mathcal{T}$, the probability $dP$ can be written as
\begin{widetext}
\begin{equation}
\label{dP_d^3k}
\begin{split}
dP=&\frac{d^3\bm{k}}{(2\pi)^3}\frac{\alpha\pi m^2}{p_-p'_-\omega}\int d\phi d\phi'\,\left\{\frac{p_-}{p'_-}+\frac{p'_-}{p_-}-4+\frac{p'_-}{p_-}\frac{\bm{p}^2_{\perp}}{m^2}-2\frac{\bm{p}_{\perp}\cdot\bm{p}'_{\perp}}{m^2}+\frac{p_-}{p'_-}\frac{\bm{p}^{\prime\,2}_{\perp}}{m^2}+\frac{k_-}{m}\left(\frac{\bm{p}_{\perp}}{p_-}-\frac{\bm{p}'_{\perp}}{p'_-}\right)\cdot[\pmb{\xi}_{\perp}(\phi)+\pmb{\xi}_{\perp}(\phi')]\right.\\
&\left.-\left[\pmb{\xi}_{\perp}^2(\phi)+\pmb{\xi}_{\perp}^2(\phi')-\left(\frac{p_-}{p'_-}+\frac{p'_-}{p_-}\right)\pmb{\xi}_{\perp}(\phi)\cdot\pmb{\xi}_{\perp}(\phi')\right]\right\}\exp\left\langle i\left\{\left(\frac{m^2+\bm{p}^{\prime 2}_{\perp}}{2p'_-}+\frac{\bm{k}_{\perp}^2}{2k_-}-\frac{m^2+\bm{p}_{\perp}^2}{2p_-}\right)(\phi-\phi')\right.\right.\\
&\left.\left.+m\int_{\phi'}^{\phi}d\tilde{\phi}\left[\frac{\bm{p}_{\perp}\cdot\pmb{\xi}_{\perp}(\tilde{\phi})}{p_-}-\frac{\bm{p}'_{\perp}\cdot\pmb{\xi}_{\perp}(\tilde{\phi})}{p'_-}+\frac{mk_-}{2p_-p'_-}\pmb{\xi}^2_{\perp}(\tilde{\phi})\right]\right\}\right\rangle,\\
=&-\frac{d^3\bm{k}}{4\pi^2}\frac{\alpha m^2}{p_-p'_-\omega}\int d\phi d\phi'\,e^{-i\frac{k_-m^2}{2p_-p'_-}\int_{\phi}^{\phi'}d\tilde{\phi}\,\left\{1+\left[\frac{\bm{p}_{\perp}}{m}-\frac{p_-}{k_-}\frac{\bm{k}_{\perp}}{m}-\bm{\xi}_{\perp}(\tilde{\phi})\right]^2\right\}}\left\{1+\frac{1}{4}\frac{p_-^2+p_-^{\prime\,2}}{p_-p'_-}[\pmb{\xi}_{\perp}(\phi)-\pmb{\xi}_{\perp}(\phi')]^2\right\},
\end{split}
\end{equation}
\end{widetext}
where $\pmb{\xi}_{\perp}(\phi)=e\bm{A}_{\perp}(\phi)/m$. This expression is particularly useful to calculate the differential probability $dP/dk_-$. In order to do this, in fact, one has to pass from the variable $k_n=\bm{k}\cdot\bm{n}$ to the variable $k_-=\sqrt{\bm{k}_{\perp}^2+k_n^2}-k_n$, i.e., to write $d^3\bm{k}=(\omega/k_-)dk_-d^2\bm{k}_{\perp}$. The integral over the emitted transverse photon momentum is Gaussian as the phase there contains at the highest quadratic terms in $\bm{k}_{\perp}$. One needs the identity (see, e.g., \cite{Gradshteyn_b_2000})
\begin{equation}
\label{G_0}
\mathcal{I}_0(a)=\int \frac{d^2\bm{z}}{(2\pi)^2}e^{ia\bm{z}^2}=\int_0^{\infty} \frac{ds}{4\pi}e^{ias}=\frac{i}{4\pi a}
\end{equation}
for any two-dimensional real vector $\bm{z}=(z_1,z_2)$ and for any constant $a$ with $\text{Im}(a)>0$. By employing this identity, one finally obtains
\begin{widetext}
\begin{equation}
\begin{split}
\frac{dP}{dk_-}=&-i\frac{\alpha}{2\pi}\frac{1}{p_-}\frac{\xi_0}{\chi_0}\int\frac{d\varphi d\varphi'}{\varphi-\varphi'+i0}\bigg\{1+\frac{p_-^2+p^{\prime\,2}_-}{4p_-p'_-}[\pmb{\xi}_{\perp}(\varphi)-\pmb{\xi}_{\perp}(\varphi')]^2\bigg\}\\
&\times\exp\left\langle i\frac{1}{2}\frac{k_-}{p'_-}\frac{\xi_0}{\chi_0}\left\{\varphi-\varphi'+\int_{\varphi'}^{\varphi}d\tilde{\varphi}\,\pmb{\xi}^2_{\perp}(\tilde{\varphi})-\frac{1}{\varphi-\varphi'}\left[\int_{\varphi'}^{\varphi}d\tilde{\varphi}\,\pmb{\xi}_{\perp}(\tilde{\varphi})\right]^2\right\}\right\rangle,
\end{split}
\end{equation}
\end{widetext}
where the laser phase $\varphi=\omega_0\phi$ ($\varphi'=\omega_0\phi'$) has been introduced, and where the prescription $\varphi-\varphi'+i0$ results from the condition on the imaginary part of the constant $a$ in Eq. (\ref{G_0}) and ensures that the emission probability vanishes if the external field vanishes. By passing now to the variables $\varphi_+=(\varphi+\varphi')/2$ and $\varphi_-=\varphi-\varphi'$ and by taking into account the parity properties of the integrand with respect to the variable $\varphi_-$, the differential probability $dP/dk_-$ is found to have the form $dP/dk_-=\int d\varphi_+\,dP/dk_-d\varphi_+$, where (see also, e.g., \cite{Di_Piazza_2017b})
\begin{widetext}
\begin{gather}
\label{f}
\frac{dP}{dk_-d\varphi_+}=\frac{\alpha}{2\pi}\frac{1}{p_-}\frac{\xi_0}{\chi_0}\text{Im}\int\frac{d\varphi_-}{\varphi_-+i0}\left\{1+\frac{p_-^2+(p_--k_-)^2}{4p_-(p_--k_-)}\left[\pmb{\xi}_{\perp}\left(\varphi_+-\frac{\varphi_-}{2}\right)-\pmb{\xi}_{\perp}\left(\varphi_++\frac{\varphi_-}{2}\right)\right]^2\right\}e^{i\Phi(k_-,\varphi_-,\varphi_+)},\\
\label{Phi}
\Phi(k_-,\varphi_-,\varphi_+)=\frac{1}{2}\frac{k_-}{p_--k_-}\frac{\xi_0}{\chi_0}\Bigg\{\varphi_-+\int_{-\varphi_-/2}^{\varphi_-/2}d\tilde{\varphi}\,\pmb{\xi}^2_{\perp}(\varphi_++\tilde{\varphi})-\frac{1}{\varphi_-}\bigg[\int_{-\varphi_-/2}^{\varphi_-/2}d\tilde{\varphi}\,\pmb{\xi}_{\perp}(\varphi_++\tilde{\varphi})\bigg]^2\Bigg\},
\end{gather}
\end{widetext}
with $\chi_0=(p_-/m)E_0/F_{cr}$. In this respect, \corr{having in mind applications based on modern numerical codes as those used in strong-field QED}, it is natural to interpret the quantity $dP/dk_-d\varphi_+$ as a probability per unit of $k_-$ and per unit of laser phase. This interpretation is consistent within the LCFA (see below and \cite{Ritus_1985}). However, we stress that if the phase $\Phi(k_-,\varphi_-,\varphi_+)$ is always of the order of unity, the radiation probability is formed only after integrating both $\varphi_+$ and $\varphi_-$ over the whole laser pulse and the above interpretation of $dP/dk_-d\varphi_+$ does not strictly hold. \cor{Finally we observe that in the limit $k_-\ll p_-$ Eqs. (\ref{f})-(\ref{Phi}) are in agreement with Eqs. (1)-(3) in \cite{Khokonov_2002}, under the approximations $k_-\approx 2\omega$ and $p_-\approx 2\varepsilon$ (see also below).}

\subsection{Photon emission probability per unit $\bm{\omega}$ and per unit $\bm{k_-}$}
It is also worth mentioning that the differential probability $dP/dk_-$ can be interpreted as (half of) the energy differential probability of the emitted photon within a very broad range of parameters. In fact, in the case of electron-laser collision this approach is justified in the most relevant configuration, which we also investigate below, where the electron is initially counterpropagating with respect to the laser field ($\bm{p}\parallel -\bm{n}$), it is ultrarelativistic ($p_-\gg m$), and its energy is much larger than $m\xi_0$ ($p_-\gg m\xi_0$, see, e.g., \cite{Di_Piazza_2014}). Under these conditions, indeed, the radiation is essentially confined within a cone along the initial momentum of the electron of maximal angular aperture of the order of $m\xi_0/p_-\ll 1$ such that $k_-\approx 2\omega$ \cite{Jackson_b_1975}. Here, we provide a more quantitative proof of the equivalence between $dP/d\omega$ and $2dP/dk_-$, and also show that it is actually violated at very low $\omega$'s, such that $\omega\lesssim\omega_0$.

In order to work on expressions of the differential probabilities which are manifestly non-negative, we notice that $\pmb{\xi}_{\perp}(\varphi)-\pmb{\xi}_{\perp}(\varphi')=\pmb{\pi}_{\perp}(\varphi')-\pmb{\pi}_{\perp}(\varphi)$, where $\pmb{\pi}_{\perp}(\varphi)=\bm{p}_{\perp}/m-(p_-/k_-)\bm{k}_{\perp}/m-\pmb{\xi}_{\perp}(\varphi)$. Thus, by integrating by parts the terms with the pre-exponent proportional to $\pmb{\pi}^2_{\perp}(\varphi)$ and to $\pmb{\pi}^2_{\perp}(\varphi')$ in the last line of Eq. (\ref{dP_d^3k}), we obtain
\begin{widetext}
\begin{equation}
\label{dP_final}
\begin{split}
dP=&\frac{\alpha}{8\pi^2}\frac{d^3\bm{k}}{p_-p'_-\omega}\frac{m^2}{\omega_0^2}\left\langle\frac{k_-^2}{p_-p'_-}\left\vert\int d\varphi\,e^{i\frac{1}{2}\frac{k_-}{p'_-}\frac{\xi_0}{\chi_0}\int_0^{\varphi}d\varphi'\,\left\{1+\left[\frac{\bm{p}_{\perp}}{m}-\frac{p_-}{k_-}\frac{\bm{k}_{\perp}}{m}-\bm{\xi}_{\perp}(\varphi')\right]^2\right\}}\right\vert^2\right.\\
&\left.+\frac{p_-^2+p_-^{\prime\,2}}{p_-p'_-}\left\vert\int d\varphi\,\left[\frac{\bm{p}_{\perp}}{m}-\frac{p_-}{k_-}\frac{\bm{k}_{\perp}}{m}-\bm{\xi}_{\perp}(\varphi)\right]e^{i\frac{1}{2}\frac{k_-}{p'_-}\frac{\xi_0}{\chi_0}\int_0^{\varphi}d\varphi'\,\left\{1+\left[\frac{\bm{p}_{\perp}}{m}-\frac{p_-}{k_-}\frac{\bm{k}_{\perp}}{m}-\bm{\xi}_{\perp}(\varphi')\right]^2\right\}}\right\vert^2\right\rangle.
\end{split}
\end{equation}
\end{widetext}
Here, it is convenient for the analytical manipulation of the integrals to regularize those in $\varphi$ which do not contain a pre-exponential factor proportional to the external field by employing the identity (see, e.g., \cite{Mackenroth_2011})
\begin{widetext}
\begin{equation}
\label{Regularization}
\begin{split}
&\int d\varphi \,e^{i\frac{1}{2}\frac{k_-}{p'_-}\frac{\xi_0}{\chi_0}\int_0^{\varphi}d\varphi'\,\{1+[\bm{s}_{\perp}+\bm{\xi}_{\perp}(\varphi')]^2\}}=-\frac{1}{1+\bm{s}^2_{\perp}} \int d\varphi \,[2\bm{s}_{\perp}\cdot\bm{\xi}_{\perp}(\varphi)+\bm{\xi}^2_{\perp}(\varphi)]e^{i\frac{1}{2}\frac{k_-}{p'_-}\frac{\xi_0}{\chi_0}\int_0^{\varphi}d\varphi'\,\{1+[\bm{s}_{\perp}+\bm{\xi}_{\perp}(\varphi')]^2\}},
\end{split}
\end{equation}
\end{widetext}
where $\bm{s}_{\perp}=(p_-/k_-)\bm{k}_{\perp}/m-\bm{p}_{\perp}/m$.

The final expression of the differential probability in Eq. (\ref{dP_final}) can now be employed to calculate either $dP/d\omega$ or $dP/dk_-$. In the first case, one simply writes $d^3\bm{k}=\omega^2d\omega d^2\bm{n}_k$, where $d^2\bm{n}_k=\sin\theta_k d\theta_k d\varphi_k$ is the differential solid angle corresponding to the direction of emission of the photon, and integrates over $d^2\bm{n}_k$, i.e.,
\begin{widetext}
\begin{equation}
\label{dP_domega}
\begin{split}
\frac{dP}{d\omega}=&\frac{\alpha}{8\pi^2}\omega\frac{m^2}{\omega_0^2}\int d^2\bm{n}_k\left\langle\frac{k_-^2}{p^2_-p^{\prime\,2}_-}\left\vert\int d\varphi\,e^{i\frac{1}{2}\frac{k_-}{p'_-}\frac{\xi_0}{\chi_0}\int_0^{\varphi}d\varphi'\,\left\{1+\left[\frac{p_-}{k_-}\frac{\bm{k}_{\perp}}{m}-\frac{\bm{p}_{\perp}}{m}+\bm{\xi}_{\perp}(\varphi')\right]^2\right\}}\right\vert^2\right.\\
&\left.+\frac{p_-^2+p_-^{\prime\,2}}{p^2_-p^{\prime\,2}_-}\left\vert\int d\varphi\,\left[\frac{p_-}{k_-}\frac{\bm{k}_{\perp}}{m}-\frac{\bm{p}_{\perp}}{m}+\bm{\xi}_{\perp}(\varphi)\right]e^{i\frac{1}{2}\frac{k_-}{p'_-}\frac{\xi_0}{\chi_0}\int_0^{\varphi}d\varphi'\,\left\{1+\left[\frac{p_-}{k_-}\frac{\bm{k}_{\perp}}{m}-\frac{\bm{p}_{\perp}}{m}+\bm{\xi}_{\perp}(\varphi')\right]^2\right\}}\right\vert^2\right\rangle.
\end{split}
\end{equation}
\end{widetext}
We call the attention to the fact that the transverse vector $\bm{r}_{\perp}=p_-\bm{k}_{\perp}/mk_-=p_-\sin\theta_k\bm{l}_{\perp}/m(1-\cos\theta_k)$, with $\bm{l}_{\perp}=(\cos\varphi_k,\sin\varphi_k)$, is effectively independent of $\omega$. Thus, it is convenient to perform the change of variable $r_{\perp}=p_-\sin\theta_k/m(1-\cos\theta_k)$ as the integral over $d^2\bm{n}_k$ can be transformed into an integral over $d^2\bm{r}_{\perp}$:
\begin{widetext}
\begin{equation}
\label{dP_domega_2}
\begin{split}
\frac{dP}{d\omega}=&\frac{\alpha}{2\pi^2}\frac{\omega}{p_-^2}\frac{\xi_0^2}{\chi_0^2}\int \frac{d^2\bm{r}_{\perp}}{(1+m^2r^2_{\perp}/p_-^2)^2}\left\langle\frac{k_-^2}{p^{\prime\,2}_-}\left\vert\int d\varphi\,e^{i\frac{1}{2}\frac{k_-}{p'_-}\frac{\xi_0}{\chi_0}\int_0^{\varphi}d\varphi'\,\{1+[\bm{r}_{\perp}-\bm{u}_{\perp}(\varphi')]^2\}}\right\vert^2\right.\\
&\left.+\frac{p_-^2+p_-^{\prime\,2}}{p^{\prime\,2}_-}\left\vert\int d\varphi\,\left[\bm{r}_{\perp}-\bm{u}_{\perp}(\varphi)\right]e^{i\frac{1}{2}\frac{k_-}{p'_-}\frac{\xi_0}{\chi_0}\int_0^{\varphi}d\varphi'\,\{1+[\bm{r}_{\perp}-\bm{u}_{\perp}(\varphi')]^2\}}\right\vert^2\right\rangle,
\end{split}
\end{equation}
\end{widetext}
where $\bm{u}_{\perp}(\varphi)=\bm{p}_{\perp}/m-\bm{\xi}_{\perp}(\varphi)$. It should be noticed that the quantities $k_-$ and $p'_-=p_--k_-$ depends on $r_{\perp}$ as $k_-=2\omega/(1+m^2r^2_{\perp}/p_-^2)$. 

In order to study now the differential probability $dP/dk_-$, we go back to Eq. (\ref{dP_final}) and we change, as explained above, from the variable $k_n$ to the variable $k_-$. Since the variables $\bm{k}_{\perp}$ and $k_-$ are then independent, it is convenient to change variable from $\bm{k}_{\perp}$ to $\bm{r}_{\perp}$ as above:
\begin{widetext}
\begin{equation}
\label{dP_dk_-}
\begin{split}
\frac{dP}{dk_-}=&\frac{\alpha}{8\pi^2}\frac{k_-}{p^2_-}\frac{\xi_0^2}{\chi_0^2}\int d^2\bm{r}_{\perp}\left\langle\frac{k_-^2}{p^{\prime\,2}_-}\left\vert\int d\varphi\,e^{i\frac{1}{2}\frac{k_-}{p'_-}\frac{\xi_0}{\chi_0}\int_0^{\varphi}d\varphi'\,\{1+[\bm{r}_{\perp}-\bm{u}_{\perp}(\varphi')]^2\}}\right\vert^2\right.\\
&\left.+\frac{p_-^2+p_-^{\prime\,2}}{p^{\prime\,2}_-}\left\vert\int d\varphi\,[\bm{r}_{\perp}-\bm{u}_{\perp}(\varphi)]e^{i\frac{1}{2}\frac{k_-}{p'_-}\frac{\xi_0}{\chi_0}\int_0^{\varphi}d\varphi'\,\{1+[\bm{r}_{\perp}-\bm{u}_{\perp}(\varphi')]^2\}}\right\vert^2\right\rangle.
\end{split}
\end{equation}
\end{widetext}
In fact, Eq. (\ref{dP_domega_2}) and Eq. (\ref{dP_dk_-}) indicate that the differential probabilities $dP/d\omega$ and $2dP/dk_-$ approximately coincide at $k_-\approx 2\omega$, if in the effective integration region in Eq. (\ref{dP_domega_2}) it is $r_{\perp}\ll p_-/m$ (which means $|\bm{k}_{\perp}|\ll k_-$). We will determine the condition when this is the case below but we can already conclude that it must be violated in the infrared limits $\omega\to 0$ and $k_-\to 0$. On the one hand, in fact, it can easily be seen from Eq. (\ref{Regularization}) and Eq. (\ref{dP_domega_2}) that $dP/d\omega$ goes linearly to zero in the limit $\omega\to 0$ as one can essentially set $\omega=0$ in the exponential functions. On the other hand, we will show below that $dP/dk_-$ tends to a constant in the limit $k_-\to 0$ (see also Appendix A). Now, an inspection to Eq. (\ref{dP_domega_2}) indicates that the condition on $r_{\perp}$ in order that  $dP/d\omega$ and $2dP/dk_-$ approximately coincide can be derived by studying the phase $\Theta(\bm{r}_{\perp},\omega,\varphi,\varphi')$ given by
\begin{widetext}
\begin{equation}
\label{Phase}
\begin{split}
\Theta(\bm{r}_{\perp},\omega,\varphi,\varphi')=&\frac{1}{2}\frac{2\omega}{(1+m^2r^2_{\perp}/p_-^2)p_--2\omega}\frac{\xi_0}{\chi_0}\int_{\varphi'}^{\varphi}d\tilde{\varphi}\,\left\{1+\left[\bm{r}_{\perp}-\frac{\bm{p}_{\perp}}{m}+\bm{\xi}_{\perp}(\tilde{\varphi})\right]^2\right\}\\
=&\frac{1}{2}\frac{2\omega}{(1+m^2r^2_{\perp}/p_-^2)p_--2\omega}\frac{\xi_0}{\chi_0}\left\{\varphi-\varphi'+\int_{\varphi'}^{\varphi}d\tilde{\varphi}\,\pmb{\xi}^2_{\perp}(\tilde{\varphi})-\frac{1}{\varphi-\varphi'}\left[\int_{\varphi'}^{\varphi}d\tilde{\varphi}\,\pmb{\xi}_{\perp}(\tilde{\varphi})\right]^2\right\}\\
&+\frac{1}{2}\frac{2\omega}{(1+m^2r^2_{\perp}/p_-^2)p_--2\omega}\frac{\xi_0}{\chi_0}(\varphi-\varphi')\left[\bm{r}_{\perp}-\frac{\bm{p}_{\perp}}{m}+\frac{1}{\varphi-\varphi'}\int_{\varphi'}^{\varphi}d\tilde{\varphi}\,\bm{\xi}_{\perp}(\tilde{\varphi})\right]^2
\end{split}
\end{equation}
\end{widetext}
and resulting from the square modulus of the integral amplitudes in Eq. (\ref{dP_domega_2}). Since we assume that $r_{\perp}$ is indeed much smaller than $p_-/m$, the term in the first line of the second equality can be ignored in our considerations. Moreover, we work in the relevant regime where $p_-\gg m$, $\chi_0\sim 1$, and $\xi_0\gg 1$, and we know already that if we look at emitted energies $\omega$ such that $2\omega\sim p_-$, then $|\varphi-\varphi'|\sim 2\pi/\xi_0$. Thus, in order to keep the phase of the order of unity, $r_{\perp}$ must be of the order of $\xi_0$ such that in the relevant region of integration the term in the square bracket in the last line in Eq. (\ref{Phase}) can undergo a compensation and become of the order of unity. This would suggest that the condition $r_{\perp}\ll p_-/m$ requires that $m\xi_0/p_-\ll 1$. However, one can argue that, since the LCFA applies here, at any phase $\varphi$ one can choose a system of coordinates aligned with the instantaneous velocity of the electron such that in that system the instantaneous transverse momentum $m\bm{u}_{\perp}(\varphi)=\bm{p}_{\perp}-m\bm{\xi}_{\perp}(\varphi)$ vanishes and the condition $r_{\perp}\ll p_-/m$ is effectively guaranteed for an ultrarelativistic electron ($m/p_-\ll 1$). The above argument does not hold in general if $2\omega\ll p_-$ and if the LCFA does not apply. In this case, since we usually consider short pulses, we can assume for the sake of definiteness that $|\varphi-\varphi'|\sim 2\pi$. By expanding the remaining terms in $1/(1+m^2r^2_{\perp}/p_-^2)$ in Eq. (\ref{dP_domega_2}) up to first order in $m^2r^2_{\perp}/p_-^2$ and by performing the integral in $r_{\perp}$, one sees that the resulting corrections can be considered small if $m\xi_0/p_-\ll 1$ and if $(p_-/2\omega)(\chi_0/\xi_0)(m^2/p_-^2)\ll 1$, i.e., if $2\omega\gg\omega_0$. The most restrictive condition is the first one, which will be however fulfilled in present and upcoming experiments in strong-field QED \cite{Di_Piazza_2014}.

\section{Results}

\subsection{Validity of the LCFA}
Our main conclusion about the validity of the LCFA can be illustrated by means of an analogy with the more familiar case of synchrotron radiation by an ultrarelativistic electron with energy $\varepsilon=m\gamma\gg m$ moving in a constant and uniform magnetic field of strength $B$ (see, e.g., the textbook \cite{Landau_b_2_1975}). By ignoring the trivial dynamics along the magnetic field, the electron moves on a circle with angular frequency $\omega_B=|e|B/m\gamma$ and the whole angular deflection undergone by the electron is $2\pi$. According to the classical theory of synchrotron radiation, the electron emits harmonics $n\omega_B$ of the fundamental angular frequency $\omega_B$ and the intensity of radiation is maximal for $n\sim\gamma^3\gg 1$, i.e., for $\omega\sim \gamma^3\omega_B\sim\chi_B\varepsilon$, where $\chi_B=\gamma B/F_{cr}$. At such large values of $n$, the intensity of the $n$th harmonic is essentially determined by combinations of Airy functions evaluated at $x_n=n^{2/3}(\gamma^{-2}+\theta^2)$, where $\theta\ll 1$ is the emission angle with respect to the instantaneous velocity of the electron. As the Airy function is exponentially suppressed for large, positive arguments, the harmonics corresponding $n\sim \gamma^3$ are mainly emitted within an angle $\theta\lesssim 1/\gamma$, much smaller than the whole angular deflection ($2\pi$) and the LCFA is applicable. However, at such low photon energies that $n\sim 1$, the ultrarelativistic limit $\gamma\gg 1$ is not sufficient for the LCFA to be applicable because radiation at large angles $\theta\sim 2\pi$ is not suppressed and contributions from the whole trajectory are virtually important. Note that the condition $n\sim 1$ corresponds to emitted photon energies $\omega\sim \chi_B\varepsilon/\gamma^3$.

Let us consider now the plane-wave case. In the regime under consideration in which $p_-\approx 2\gamma m\gg m$, and $p_-\gg m\xi_0$, the whole angular deflection undergone by the electron is of the order of $m\xi_0/p_-$, whereas the instantaneous emission still occurs along the velocity within a cone of angular aperture $\sim 1/\gamma\approx 2m/p_-$. Since, analogously to the synchrotron case, the intensity of radiation is maximal for photon light-cone energies $k_-\sim\chi_0p_-$, the condition $\xi_0\gg 1$ is sufficient to guarantee that for these values of $k_-$ the LCFA applies. Thus, the parameter $\xi_0$ plays the role of $\gamma$ in the synchrotron case and, in analogy with the latter, we expect that for such small light-cone energies $k_-\lesssim \chi_0p_-/\xi_0^3$, the LCFA fails even if $\xi_0\gg 1$.

The validity of this condition can indeed be shown in a more quantitative and rigorous way by analyzing the structure of the phase in Eq. (\ref{Phi}). In fact, if the integral in $\varphi_-$ in Eq. (\ref{f}) is formed within a region $|\varphi_-|\lesssim\varphi_f/2$ and $\varphi_f$ (the laser formation phase) is much smaller than, say, $2\pi$, then we can expand the phase $\Phi(k_-,\varphi_-,\varphi_+)$ as
\begin{equation}
\Phi(k_-,\varphi_-,\varphi_+)\approx\frac{1}{2}\frac{\xi_0}{\rho_0(k_-)}\left[\varphi_-+\frac{1}{12}\pmb{\xi}^{\prime\,2}_{\perp}(\varphi_+)\varphi_-^3\right],
\end{equation}
where $\rho_0(k_-)=\chi_0(p_--k_-)/k_-$. By identifying the formation phase $\varphi_f$ as the value of $\varphi_-$ such that $|\Phi(k_-,\varphi_f/2,\varphi_+)-\Phi(k_-,-\varphi_f/2,\varphi_+)|=\pi$ and by assuming that $|\pmb{\xi}'_{\perp}(\varphi_+)|\sim \xi_0$, we obtain
\begin{equation}
\label{FL_xi_large_rho_unity}
\begin{split}
\varphi_f&=\frac{8}{|\pmb{\xi}'_{\perp}(\varphi_+)|}\sinh\left(\frac{1}{3}\sinh^{-1}\left(\frac{3\pi}{4}\frac{\rho_0(k_-)}{\xi_0}|\pmb{\xi}'_{\perp}(\varphi_+)|\right)\right)\\
&\sim\frac{8}{\xi_0}\sinh\left(\frac{1}{3}\sinh^{-1}\left(\frac{3\pi}{4}\rho_0(k_-)\right)\right).
\end{split}
\end{equation}
In the nonlinear regime $\xi_0\gg 1$, the laser formation phase is much smaller than $2\pi$ for $\rho_0(k_-)\lesssim 1$. However, for sufficiently small values of $k_-$ for a given value of $\chi_0$, the quantity $\rho_0(k_-)$ can become so large that the LCFA can be invalidated, because for $\rho_0(k_-)\gg 1$, then $\varphi_f\sim 4(3\pi/2)^{1/3}\rho^{1/3}_0(k_-)/\xi_0$. Indeed, the parametric condition for the validity of the LCFA coincides with the one given above in analogy with the synchrotron case. Correspondingly, we show below that the corrections to the laser formation phase in Eq. (\ref{FL_xi_large_rho_unity}) scale with the square of the parameter $\rho^{1/3}_0(k_-)/\xi_0$. We have also seen that if one looks at sufficiently small values of $k_-$ for a given value of $\chi_0$, the quantity $\rho_0(k_-)$ can become so large that the LCFA can be invalidated. We use the expression of $\varphi_f$ in this limit [$\varphi_f\sim 4(3\pi/2)^{1/3}\rho^{1/3}_0(k_-)/\xi_0$] to determine the correction to the laser formation phase arising from higher-order terms in the expansion of the function $\pmb{\xi}_{\perp}(\varphi_++\tilde{\varphi})$ around $\tilde{\varphi}=0$. Now, the next term $\delta\Phi(k_-,\varphi_-,\varphi_+)$ in the mentioned expansion is given by (see also \cite{Baier_1989,Khokonov_2002})
\begin{equation}
\delta\Phi(k_-,\varphi_-,\varphi_+)=\xi_0\frac{\pmb{\xi}^{\prime\prime\,2}_{\perp}(\varphi_+)+3\pmb{\xi}^{\prime}_{\perp}(\varphi_+)\cdot\pmb{\xi}^{\prime\prime\prime}_{\perp}(\varphi_+)}{\rho_0(k_-)}\frac{\varphi_-^5}{1440}.
\end{equation}
By defining the formation length as in the main text with the expression of $\Phi(k_-,\varphi_-,\varphi_+)$ expanded up to $\varphi_-^5$ and by estimating $|\pmb{\xi}^{\prime\prime\,2}_{\perp}(\varphi_+)+3\pmb{\xi}^{\prime}_{\perp}(\varphi_+)\cdot\pmb{\xi}^{\prime\prime\prime}_{\perp}(\varphi_+)|\sim \xi_0^2$, we obtain that the absolute value of the correction $\delta\varphi_f$ to the phase formation length at $1\ll \rho_0(k_-)\ll \xi_0^3$ scales as
\begin{align}
|\delta\varphi_f|\sim\frac{\rho_0(k_-)}{\xi^3_0}&& \Longrightarrow &&\frac{|\delta\varphi_f|}{\varphi_f}\sim\frac{\rho^{2/3}_0(k_-)}{\xi^2_0}\ll 1.
\end{align}
The corrections arising both in the phase and in the pre-exponential function in Eq. (\ref{f}) show the same scaling such that we expect substantial deviations from the LCFA for $\rho^{2/3}_0(k_-)\gtrsim\xi^2_0$, i.e., for $k_-/p_-\lesssim \chi_0/\xi_0^3$, which indeed is the parameter determining the scaling of the field-dependent terms in the phase $\Phi(k_-,\varphi_-,\varphi_+)$ [see Eq. (\ref{Phi})].

\subsection{Numerical evaluation: exact results vs LCFA}

Let us consider now a numerical example in which a linearly polarized plane wave with $\omega_0=1.55\;\text{eV}$ and peak intensity $I_0=E_0^2/4\pi=4.4\times 10^{20}\;\text{W/cm$^2$}$ ($\xi_0\approx 10$), and an electron initially counterpropagating with respect to the plane wave with $\varepsilon=10\;\text{GeV}$ ($\chi_0\approx 1.2$). The laser pulse shape is chosen in such a way that if $\bm{A}_{\perp}(\varphi)=\bm{A}_{\perp,0}\psi(\varphi)$, with $\bm{n}\cdot\bm{A}_{\perp,0}=0$ and $|\bm{A}_{\perp,0}|=E_0/\omega_0$, then $\psi(\varphi)=\exp(-\varphi^2/\Delta\varphi^2)\sin(\varphi+\varphi_0)$. Here, $\varphi_0$ is the carrier envelope phase and the width $\Delta\varphi$ is related to the full width half maximum (FWHM) of the intensity by the relation $\Delta\varphi=\text{FWHM}/\sqrt{2\log 2}$. We have set $\varphi_0=\pi/2$ and the FWHM corresponding to $5\;\text{fs}$. Since in this case the approximation $k_-\approx 2\omega$ is valid for $\omega\gg\omega_0$ (see the Appendix A), the photon probability $2dP/dk_-\approx dP/d\omega$ in units of $1/m$ as a function of $k_-/p_-$ is shown in Fig.~1, where the solid red curve corresponds to the exact calculation from Eq. (\ref{f}) and the dotted black curve to the one within the LCFA, which we denote by $dP_{\text{LCFA}}/dk_-$ (see, e.g., \cite{Baier_b_1998}).
\begin{figure}
\begin{center}
\includegraphics[width=\columnwidth]{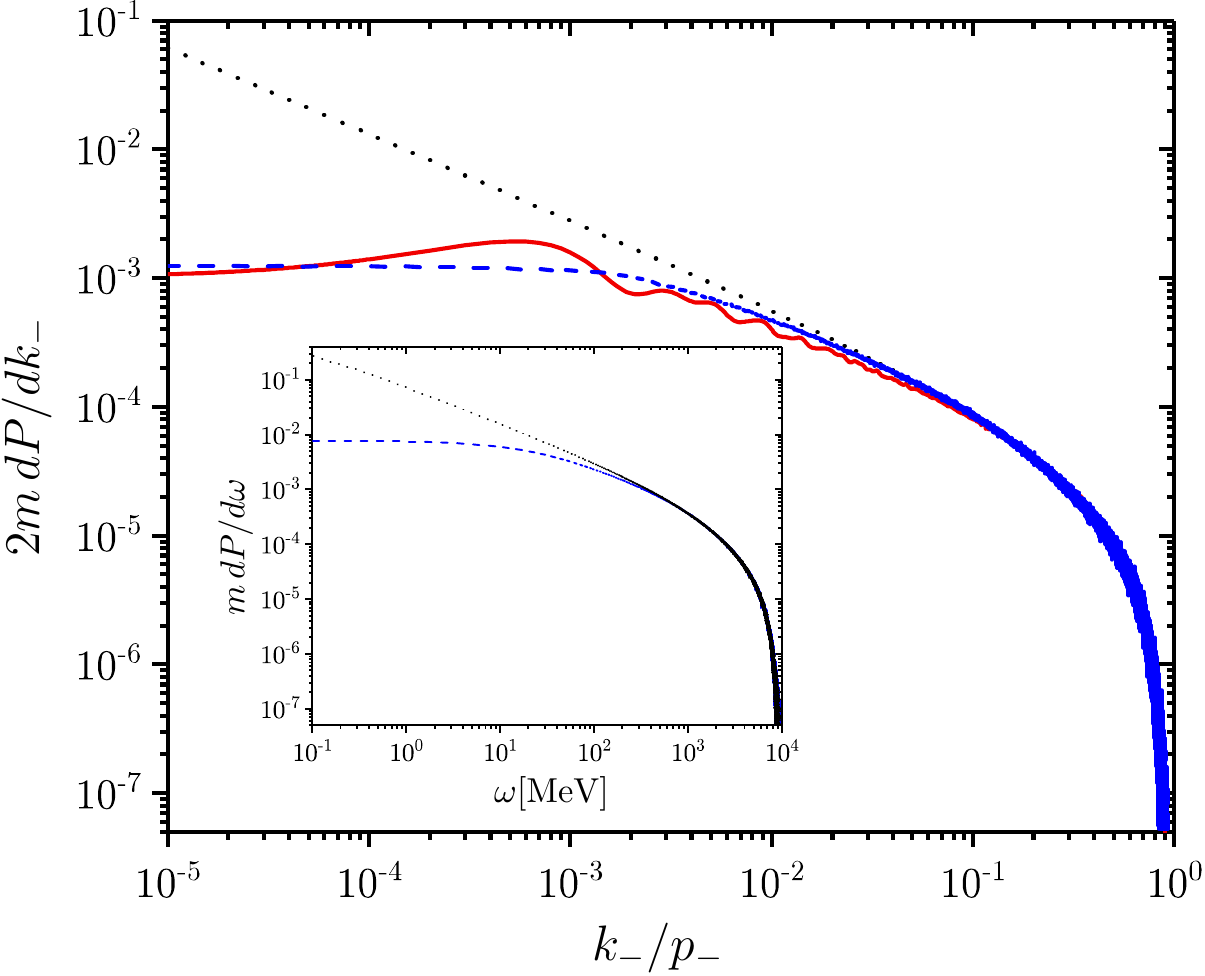}
\caption{Exact (solid red curve) vs local constant field approximated (dotted black curve) differential photon emission probability for an electron with initial energy of $10\;\text{GeV}$ colliding head-on with a plane wave pulse of $5\;\text{fs}$ FWHM duration and $4.4\times 10^{20}\;\text{W/cm$^2$}$ peak intensity. The dashed blue curve shows the same probability obtained via the numerical code presented in \cite{Tamburini_2017}, with the improved emission model as described in the text. The inset shows the corresponding probabilities with the same color code and calculated via the numerical code in \cite{Tamburini_2017} in the case of an electron beam with $10\;\text{GeV}$ average energy and $10\,\%$ energy spread colliding head-on with a focused Gaussian laser beam with $30\;\text{fs}$ FWHM duration, $4.4\times 10^{20}\;\text{W/cm$^2$}$ peak intensity and $8\;\text{$\mu$m}$ waist radius.}
\end{center}
\end{figure}

The figure clearly shows that although $\xi_0\gg 1$ the LCFA fails in the infrared region where it predicts a quantitatively and qualitatively different behavior of the photon probability with respect to the exact result. In fact, the LCFA predicts that the differential photon probability diverges as $(p_-/k_-)^{2/3}$ in the limit $k_-/p_-\to 0$ (see, e.g., \cite{Baier_b_1998}), whereas the exact differential photon probability approaches the constant value
\begin{equation}
\left.\frac{dP}{dk_-}\right|_0=\lim_{k_-\to 0}\frac{dP}{dk_-}=\frac{\alpha}{2}\frac{1}{p_-}\frac{\xi_0}{\chi_0}\int d\varphi\,\pmb{\xi}^2_{\perp}(\varphi)
\end{equation}
in the same limit. This result is derived analytically in the Appendix A. However, since $\lim_{k_-\to 0}\Phi(k_-,\varphi_-,\varphi_+)=0$ [see Eq. (\ref{Phi})] we expect that nonlinear effects are suppressed for $k_-\to 0$ (see also \cite{Meuren_2016}), which results in $dP/dk_-|_0$ being proportional to $\xi_0^2$. Thus, we have been able to derive this result also by starting from the differential cross section $d\sigma/dk_-$ of linear Compton scattering [see, e.g., Eq. (86.6) in \cite{Landau_b_4_1982}]. In fact, in our notation and in the limit of vanishing photon recoil ($k_-\ll p_-$) this quantity reads $\lim_{k_-\to 0}d\sigma/dk_-=d\sigma/dk_-|_0=2\pi\alpha^2/\omega_lp^2_-$, where we have assumed that the incoming (laser) photon has angular frequency $\omega_l$. Now, by introducing the Fourier transform $\tilde{\pmb{A}}_{\perp}(\omega_l)=\int d\phi\, \pmb{A}_{\perp}(\varphi)e^{i\omega_l\phi}$, the number $dN_l/d\Sigma d\omega_l$ of incoming laser photons per unit surface $\Sigma$ and unit frequency $\omega$ is given by $(4\pi^2)^{-1}\omega_l|\tilde{\pmb{A}}_{\perp}(\omega_l)|^2$. Thus, the above given asymptotic value of the photon probability is obtained by multiplying $d\sigma/dk_-|_0$ by $dN_l/d\Sigma d\omega_l$, by integrating with respect to $\omega_l$ and by exploiting the Parseval identity. 

Note that the probability $\Delta P$ of emitting a photon with $k_-$ between zero and $\Delta k_-$ tends to zero for $\Delta k_-\to 0$. For the numerical parameters employed in Fig.~1, it is $2mdP/dk_-|_0\approx 1.01\times 10^{-3}$ in excellent agreement with the numerical value $1.05\times 10^{-3}$ at $k_-=5\times 10^{-6}p_-$. As a check of the analytical predictions of the values of $k_-$ below which the LCFA fails, one can verify that at $k_-/p_-\approx 3\times 10^{-3}$ and for the above mentioned parameters, the correction to the LCFA is expected to scale as the parameter $(\chi_0p_-/k_-)^{2/3}/\xi_0^2\approx 0.5$, which coincides with the relative difference $2|dP/dk_--dP_{\text{LCFA}}/dk_-|/(dP/dk_-+dP_{\text{LCFA}}/dk_-)\approx 0.5$. We also stress that, although photons with $k_-\approx 3\times 10^{-3}\,p_-$ are identified as ``infrared'', they have an energy of $\omega\approx k_-/2=30\;\text{MeV}$. \cor{However, since the differences between the exact spectrum and the one calculated within the LCFA arise for photon energies much smaller than the electron energy, we expect that the effects investigated here will not have a dramatic impact on the dynamics of the emitting electron, except as a cumulative effect in the case of several emissions. Nevertheless, the photon spectra, which also represent an important physical observable, show substantial differences if the exact formulas are employed instead of the LCFA.} In this respect, we point out as a final remark that the total probability of emitting a photon is $0.69$ according to the exact QED expression from Eq. (\ref{f}), which is significantly smaller than the value of $0.93$ predicted by the LCFA.

At this point, it is also interesting to study the infrared behavior of the differential probability per unit phase $dP/dk_-d\varphi_+$, because this is precisely the quantity employed in numerical codes, and the difference between the LCFA and the exact theory is even more striking than for the integrated quantities investigated above. In fact, the double differential probability $dP_{\text{LCFA}}/dk_-d\varphi_+$ within the LCFA shows the same asymptotic behavior $\sim (p_-/k_-)^{2/3}$ in the infrared region as $dP_{\text{LCFA}}/dk_-=\int d\varphi_+\,dP_{\text{LCFA}}/dk_-d\varphi_+$, because $dP_{\text{LCFA}}/dk_-d\varphi_+$ is proportional to $\chi^{2/3}(\varphi_+)$ in that region, which is integrable in $\varphi_+$ for a pulsed plane wave. In sharp contrast, the general expression $dP/dk_-d\varphi_+$ in Eq.~(\ref{f}), with the limitations indicated below Eq. (\ref{Phi}), tends to zero as \corr{$(k_-/p_-)[\log(p_-/k_-)+b]$} in the limit $k_-\to 0$, with the quantity $b$ being independent of $k_-$ (see Appendix A).

\subsection{Improved LCFA}

Having elucidated the shortcomings of the LCFA at low photon energies, we describe now a possible scheme to implement the probability of nonlinear Compton scattering beyond the LCFA in advanced numerical codes aiming at describing laser-matter interaction including quantum effects. The method is based on the above remark that at very low light-cone energies, nonlinear effects become less important such that the probability of linear Compton scattering is expected to work reasonably well. This probability can be easily derived from the cross section of linear Compton scattering given in Eq. (86.6) in \cite{Landau_b_4_1982}. However, only photons with $k_-\le k_{-,\text{LCS}}=p_-/(1+\xi_0/2\chi_0)$ can be emitted via linear Compton scattering \cite{Landau_b_4_1982}. Also, according to the above findings, we decide to use the LCFA for $k_->k_{-,\text{LCFA}}$, where $k_{-,\text{LCFA}}$ is such that the formation length $\varphi_f$ in Eq. (\ref{FL_xi_large_rho_unity}) is, for the sake of definiteness, equal to $2\pi$, which gives
\begin{equation}
k_{-,\text{LCFA}}=\frac{p_-}{1+\frac{4}{3\pi\chi_0}\sinh\left(3\sinh^{-1}\left(\frac{\pi}{4}\xi_0\right)\right)}.
\end{equation}
Since it is $k_{-,\text{LCFA}}\le k_{-,\text{LCS}}$, at each space-time point the numerical code evaluates the local values of $k_{-,\text{LCS}}$ and $k_{-,\text{LCFA}}$ and uses the probability of linear Compton scattering for $k_-\le k_{-,\text{LCFA}}$ and the LCFA for $k_-> k_{-,\text{LCFA}}(\varphi_+)$. It is worth pointing out that the employed local photon emission probability per unit $k_-$ and unit $\varphi_+$ exhibits a discontinuity at $k_{-,\text{LCFA}}(\varphi_+)$. However, \cor{the physical origin of this discontinuity is clear, as the} LCFA takes into account the absorption of an arbitrary number of laser photons, whereas the linear theory only includes the absorption of a single laser photon. Moreover, the integrated probability over $k_-$, which is used in the code, is continuous like the resulting emission spectrum (see Fig.~1).

We have implemented this method in the multiparticle quantum code presented in \cite{Tamburini_2017}, which describes the interaction of electrons, positrons and photons with an intense laser field of arbitrary space-time structure including the two basic quantum processes: nonlinear Compton scattering and nonlinear Breit-Wheeler pair production. As a benchmark, the results above for the numerical example are reported in Fig. 1 (dashed blue curve). The figure shows good agreement between the exact red curve and the implemented model. In particular, the latter predicts a total probability of $0.71$ while the exact value is $0.69$. \cor{We also note that our model cannot reproduce the oscillations shown by the red curve in Fig. 1, which arise from the interference among the amplitudes corresponding to the absorption of different number of laser photons}. Other two benchmark examples are shown in the Appendix B. 

The results of the simulation of a more realistic situation, where \cor{the laser beam is also spatially focused and a bunch of electrons with a given initial spatial and momentum distribution collides with the laser field, is shown in the inset of Fig. 1. Our main aim concerning this example is to show that our method can also be implemented in situations where} each electron on average emits more than one photon. The linearly polarized laser field is now modeled by a Gaussian field with waist size $8\;\text{$\mu$m}$ and pulse duration $30\;\text{fs}$, with the other parameters coinciding with those in the plane-wave case. Moreover, a beam of $10^8$ electrons has been simulated with transverse diameter of $8\;\text{$\mu$m}$, length of $10\;\text{$\mu$m}$, Gaussian energy distribution centered at $10\;\text{GeV}$, $10\,\%$ energy spread, and $1\;\text{mrad}$ angular aperture. The results have been obtained with the code described in \cite{Tamburini_2017}, with the probability of photon emission given either by the LCFA (dotted black curve) or by the improved method described above (dashed blue curve). The inset clearly shows the significant difference in the low-energy part of the photon spectra as compared to the results obtained by employing exclusively the LCFA. The total number of photons after the interaction according to the LCFA is $4.4\times 10^8$, whereas our improved model predicts $3.1\times 10^8$ produced photons.

\section{Conclusions}
In conclusion, we have found in the case of non-linear Compton scattering that even for a plane wave such that the condition $\xi_0\gg 1$ is fulfilled, the local constant field approximation predicts a photon yield, which qualitatively and quantitatively differs from the exact one at sufficiently low emitted photon energies. Since numerical codes exclusively employ emission probabilities within the local constant field approximation, we have worked out an improved implementation of the (differential) photon emission probability, which remedies the shortcomings of the local constant field approximation in the low-energy region of the spectrum. Our numerical simulations indicate that the studied effects and differences can be measured in principle already with available technology.

\begin{acknowledgments}
We acknowledge useful discussions with A. Angioi, S. Bragin, N. J. Fisch, K. Z. Hatsagortsyan, M. E. Peskin, and T. N. Wistisen. SM was supported by the German Research Foundation (Deutsche Forschungsgemeinschaft, DFG) -- ME 4944/1-1.
\end{acknowledgments}

\appendix

\section{Asymptotic behavior of the photon emission probability for $\bm{k_-}\bm{\to} \bm{0}$}

In this appendix we derive the asymptotic behavior of $dP/dk_-d\varphi_+$ and $dP/dk_-$ in the limit $k_-\to 0$.
In the first case, we can conveniently write the probability $dP/dk_-d\varphi_+$ as the sum of the two terms $dP_0/dk_-d\varphi_-$ and $dP_f/dk_-d\varphi_+$, with the pre-exponential function of the term $dP_0/dk_-d\varphi_+$ ($dP_f/dk_-d\varphi_+$) being independent of (dependent on) the external field [see Eq. (\ref{f})]. Now, the term $dP_0/dk_-d\varphi_+$ can be written as 
\begin{equation}
\label{f_0}
\frac{dP_0}{dk_-d\varphi_+}=\frac{\alpha}{\pi}\frac{1}{p_-}\frac{\xi_0}{\chi_0}\text{Im}\int_0^{\infty}\frac{d\varphi_-}{\varphi_-}\left[e^{ig\Psi(\varphi_-,\varphi_+)}-e^{ig\varphi_-}\right],
\end{equation}
where $g=(1/2)(k_-/p'_-)(\xi_0/\chi_0)$ and where $\Psi(\varphi_-,\varphi_+)=\Phi(k_-,\varphi_-,\varphi_+)/g\equiv\varphi_-+\Psi_f(\varphi_-,\varphi_+)$. The asymptotic behavior for $k_-\to 0$ (i.e., $g\to 0$) is found by splitting the integral into two regions by introducing an intermediate scale $\varphi_-^*$ such that $1\ll\varphi_-^*\ll 1/g$. In the final result the intermediate scale $\varphi_-^*$ drops out and one easily finds that
\begin{equation}
\label{dP_0_dk_-}
\begin{split}
&\left.\frac{dP_0}{dk_-d\varphi_+}\right\vert_{k_-\to 0} \sim\frac{\alpha}{2\pi}\frac{k_-}{p_-^2}\frac{\xi^2_0}{\chi^2_0}\left\{\left[\log\left(\frac{2\chi_0}{\xi_0}\frac{p_-}{k_-}\right)-C\right]\right.\\
&\qquad\left.\times\Psi_f(\infty,\varphi_+)-\int_0^{\infty}d\varphi_-\log(\varphi_-)\frac{\partial\Psi_f(\varphi_-,\varphi_+)}{\partial\varphi_-}\right\}, 
\end{split}
\end{equation}
where $C=0.577...$ is the Euler constant. The asymptotic behavior of the remaining term $dP_f/dk_-d\varphi_+$ is easily obtained because the external field in the pre-exponent ensures the convergence of the integral in $\varphi_-$:
\begin{equation}
\label{dP_f_dk_-}
\begin{split}
&\left.\frac{dP_f}{dk_-d\varphi_+}\right\vert_{k_-\to 0}\sim\frac{\alpha}{4\pi}\frac{k_-}{p_-^2}\frac{\xi^2_0}{\chi^2_0}\int_0^{\infty}\frac{d\varphi_-}{\varphi_-}\Psi(\varphi_-,\varphi_+)\\
&\qquad\times\left[\pmb{\xi}_{\perp}\left(\varphi_+-\frac{\varphi_-}{2}\right)-\pmb{\xi}_{\perp}\left(\varphi_++\frac{\varphi_-}{2}\right)\right]^2.
\end{split}
\end{equation}

We pass now to the asymptotic behavior of $dP/dk_-$ in the same limit $k_-\to 0$. For the sake of definiteness, we assume that the laser field is linearly polarized along the $x$ direction. Thus, Eq. (\ref{dP_dk_-}) can be written in the form
\begin{widetext}
\begin{equation}
\begin{split}
\frac{dP}{dk_-}=&\frac{\alpha}{8\pi^2}\frac{k_-}{p_-p'_-}\frac{\xi_0^4}{\chi_0^2}\int d\varphi d\varphi'\,\left\{\psi(\varphi)\psi(\varphi')\left[\frac{p_-^2+p_-^{\prime\,2}}{p_-p'_-}I_{0,0}(k_-,\varphi,\varphi')-8I_{2,2}(k_-,\varphi,\varphi')\right]\right.\\
&\quad+\xi_0[\psi(\varphi)\psi^2(\varphi')+\psi(\varphi')\psi^2(\varphi)]\left[\frac{p_-^2+p_-^{\prime\,2}}{p_-p'_-}I_{1,1}(k_-,\varphi,\varphi')-4I_{1,2}(k_-,\varphi,\varphi')\right]\\
&\left.\quad+\xi_0^2\psi^2(\varphi)\psi^2(\varphi')\left[\frac{p_-^2+p_-^{\prime\,2}}{p_-p'_-}I_{0,1}(k_-,\varphi,\varphi')-2I_{0,2}(k_-,\varphi,\varphi')\right]\right\},
\end{split}
\end{equation}
\end{widetext}
where we have introduced the integrals
\begin{equation}
\label{I_jk}
\begin{split}
&I_{j,k}(k_-,\varphi,\varphi')=\text{Re}\int d^2\bm{s}_{\perp}\frac{s_x^j}{(1+\bm{s}^2_{\perp})^k}\\
&\quad\times e^{i\frac{1}{2}\frac{k_-}{p'_-}\frac{\xi_0}{\chi_0}\int_{\varphi'}^{\varphi}d\tilde{\varphi}\,\{1+[\bm{s}_{\perp}+\bm{\xi}_{\perp}(\tilde{\varphi})]^2\}},
\end{split}
\end{equation}
with $j,k=0,1,2$, with respect to the already introduced dimensionless transverse vector $\bm{s}_{\perp}=(p_-/k_-)\bm{k}_{\perp}/m-\bm{p}_{\perp}/m$. A straightforward power-law analysis of the integrand in Eq. (\ref{I_jk}) shows that both $I_{0,2}(k_-,\varphi,\varphi')$ and $I_{1,2}(k_-,\varphi,\varphi')$ are finite in the limit $k_-\to 0$ ($I_{0,2}(0,\varphi,\varphi')=\pi$ and $I_{1,2}(0,\varphi,\varphi')=0$), whereas $I_{0,1}(k_-,\varphi,\varphi')$ and $I_{2,2}(k_-,\varphi,\varphi')$ diverge logarithmically in the same limit. The analysis of the remaining integrals $I_{0,0}(k_-,\varphi,\varphi')$ and $I_{1,1}(k_-,\varphi,\varphi')$ is more involved and we present here some details. Starting from the integral $I_{1,1}(k_-,\varphi,\varphi')$, a straightforward power-law analysis of it would indicate a linear divergence in the limit $k_-\to 0$. However, one can qualitatively observe that the leading term in $k_-\to 0$ vanishes due to the symmetry properties of the integrand (the integral is rigorously speaking divergent). More quantitatively by performing the integral in the angular polar coordinate, one obtains
\begin{widetext}
\begin{equation}
\begin{split}
I_{1,1}(k_-,\varphi,\varphi')=&-\pi\eta(\varphi,\varphi')\text{Re}\,e^{i\tilde{\Phi}(k_-,\varphi,\varphi')}\int_0^{\infty} dr \frac{1}{\sqrt{1+r^2+2rb(\varphi,\varphi')}}\\
&\times \left[1-\frac{2r}{1+r+\sqrt{1+r^2+2rb(\varphi,\varphi')}}\right]e^{i\frac{1}{2}\frac{k_-}{p'_-}\frac{\xi_0}{\chi_0}\varphi_-[1+\eta^2(\varphi,\varphi')]r},
\end{split}
\end{equation}
\end{widetext}
where $\eta(\varphi,\varphi')=(\varphi-\varphi')^{-1}\int_{\varphi'}^{\varphi}d\tilde{\varphi}\,\psi(\tilde{\varphi})$, $\tilde{\Phi}(k_-,\varphi,\varphi')=\Phi(k_-,\varphi-\varphi',(\varphi+\varphi')/2)$ (see Eq. (\ref{Phi})), and $b(\varphi,\varphi')=[1-\eta^2(\varphi,\varphi')]/[1+\eta^2(\varphi,\varphi')]$. From this expression one can easily show that $I_{1,1}(0,\varphi,\varphi')=-\pi\eta(\varphi,\varphi')$. Finally, we consider the integral $I_{0,0}(k_-,\varphi,\varphi')$ and we observe that rigorously speaking it is not convergent. For this reason, in order to obtain its correct value in the limit $k_-\to 0$, it is easier to return to the original order of integration, where the integral in $\bm{s}_{\perp}$ is performed after the integral in $\varphi$. Since the presence of the functions $\psi(\varphi)$ and $\psi(\varphi')$ in the pre-exponent limits the effective range of the variables $\varphi$ and $\varphi'$, we can write that
\begin{widetext}
\begin{equation}
\lim_{k_-\to 0}\text{Re}\int d^2\bm{s}_{\perp}\int d\varphi d\varphi'\,\psi(\varphi)\psi(\varphi')e^{i\frac{1}{2}\frac{k_-}{p'_-}\frac{\xi_0}{\chi_0}\int_{\varphi'}^{\varphi}d\tilde{\varphi}\,\{1+[\bm{s}_{\perp}+\bm{\xi}_{\perp}(\tilde{\varphi})]^2\}}\approx\lim_{k_-\to 0}\int d^2\bm{s}_{\perp}\left\vert\int d\varphi \,\psi(\varphi)e^{i\frac{1}{2}\frac{k_-}{p_-}\frac{\xi_0}{\chi_0}\varphi s_{\perp}^2}\right\vert^2.
\end{equation}
\end{widetext}
By appropriately rescaling the variable $\bm{s}_{\perp}$, it is clear that this integral turns to be inversely proportional to $k_-$. Thus, this term provides the leading contribution to $dP/dk_-$ in the limit $k_-\to 0$ and we have that
\begin{equation}
\begin{split}
\lim_{k_-\to 0}\frac{dP}{dk_-}&=\frac{\alpha}{2\pi}\frac{1}{p_-}\frac{\xi_0}{\chi_0}\int_0^{\infty}d\rho\,|\tilde{\pmb{\xi}}_{\perp}(\rho)|^2,
\end{split}
\end{equation}
where $\tilde{\pmb{\xi}}_{\perp}(\rho)=\int d\varphi\, \pmb{\xi}_{\perp}(\varphi)e^{i\rho\varphi}$ is the Fourier transform of the field $\pmb{\xi}_{\perp}(\varphi)$. By using the Parseval identity, one finally obtains
\begin{equation}
\label{const}
\lim_{k_-\to 0}\frac{dP}{dk_-}=\frac{\alpha}{2}\frac{1}{p_-}\frac{\xi_0}{\chi_0}\int d\varphi\,\pmb{\xi}^2_{\perp}(\varphi).
\end{equation}
Going back to Eq. (\ref{dP_dk_-}), we notice that it is crucial that the integral in $\bm{r}_{\perp}$ is performed over the whole plane, i.e., that the limit $r_{\perp}\to\infty$ is performed before the limit $k_-\to 0$. In fact, the two limits do not commute, which is not surprising because the definition of $\bm{r}_{\perp}$ includes $k_-$. Thus, if we perform the integral in $r_{\perp}$ up to a fixed $R_{\perp}$, the resulting probability would tend to zero (it is also evident that $\lim_{k_-\to 0} dP/dk_-d^2\bm{r }_{\perp}=0$).

\section{Benchmark examples of the numerical implementation}
Below we provide two numerical examples as benchmark of the photon emission probability beyond the LCFA described in the main text. 

In the first example we have considered the same numerical parameters as in the main text except that the laser full width half maximum (FWHM) of the intensity is $2.5\;\text{fs}$ (see Fig. 2). The integrated probability is $0.34$ according to the exact expression of the probability from Eq. (\ref{f}) and $0.36$ according to the improved emission model with respect to the LCFA, whereas the latter predicts $0.46$.
\begin{figure}
\label{Benchmark_1}
\begin{center}
\includegraphics[width=\columnwidth]{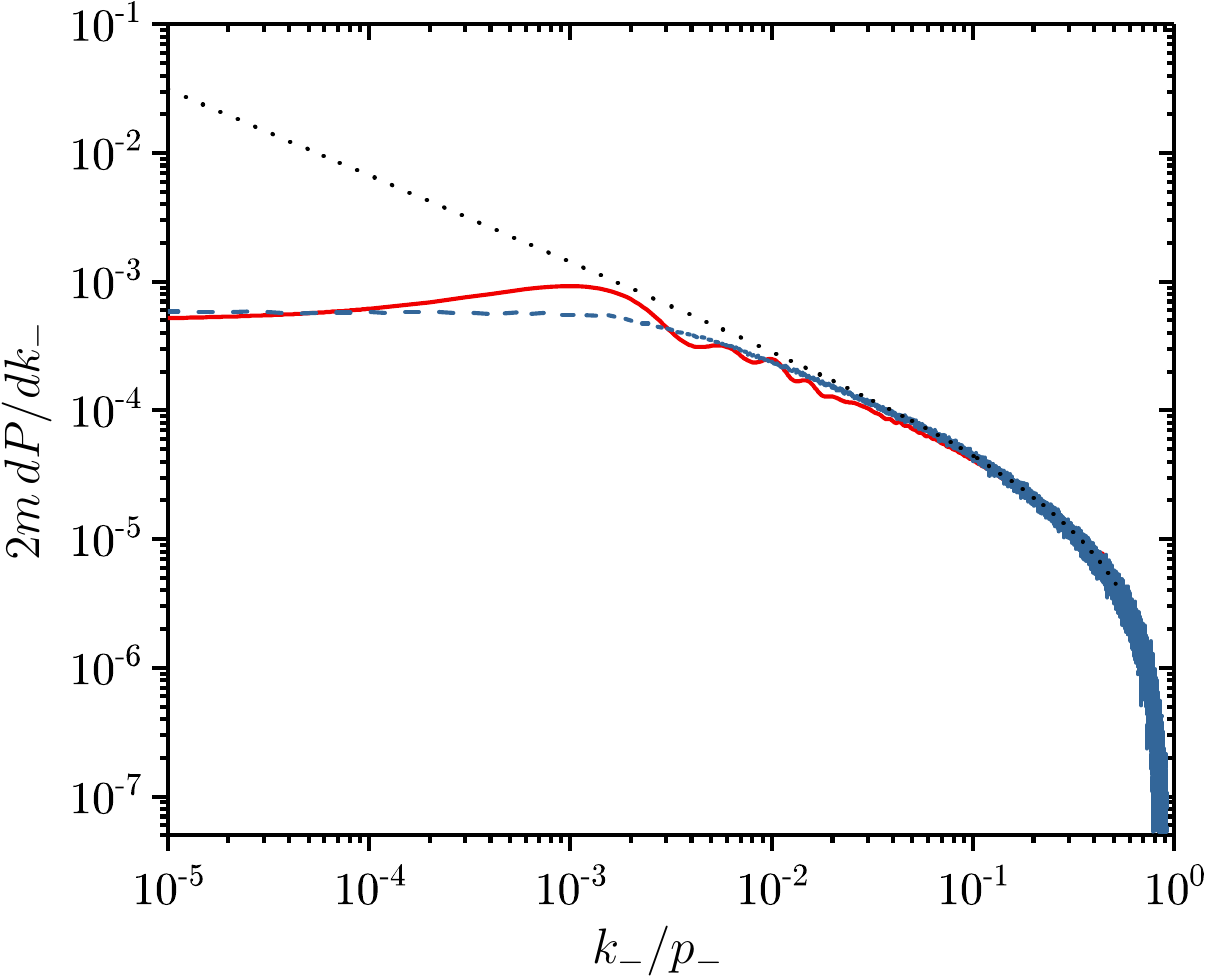}
\caption{\corr{Exact (solid red curve) vs local constant field approximated (dotted black curve) differential photon emission probability for an electron with initial energy of $10\;\text{GeV}$ colliding head-on with a plane wave pulse of $2.5\;\text{fs}$ FWHM duration and $4.4\times 10^{20}\;\text{W/cm$^2$}$ peak intensity. The dashed blue curve shows the same probability obtained via the numerical code presented in \cite{Tamburini_2017}, with the improved emission model as described in the main text.}}
\end{center}
\end{figure}

In the second example we have considered the same numerical parameters as in the main text except that the laser intensity is $2.7\times 10^{20}\;\text{W/cm$^2$}$, corresponding to $\xi_0=8$ (see Fig. 3). The integrated probability is $0.50$ according to the exact expression of the probability from Eq. (\ref{f}) and $0.54$ according to the improved emission model with respect to the LCFA, whereas the latter predicts $0.72$.
\begin{figure}
\label{Benchmark_2}
\begin{center}
\includegraphics[width=\columnwidth]{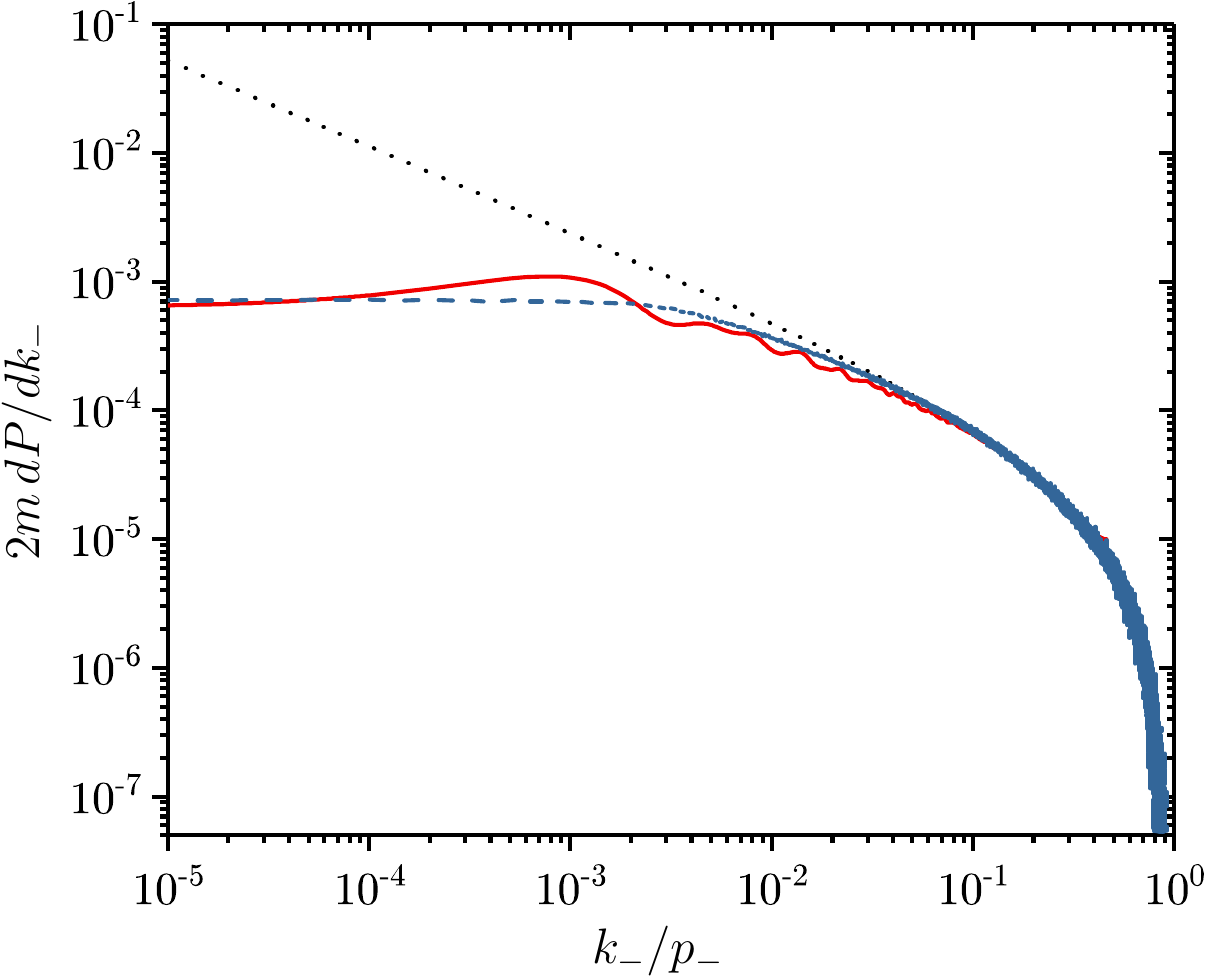}
\caption{\corr{Exact (solid red curve) vs local constant field approximated (dotted black curve) differential photon emission probability for an electron with initial energy of $10\;\text{GeV}$ colliding head-on with a plane wave pulse of $5\;\text{fs}$ FWHM duration and $2.7\times 10^{20}\;\text{W/cm$^2$}$ peak intensity. The dashed blue curve shows the same probability obtained via the numerical code presented in \cite{Tamburini_2017}, with the improved emission model as described in the main text.}}
\end{center}
\end{figure}

%


\end{document}